\documentstyle[preprint,eqsecnum,aps,12pt]{revtex}
\begin{document}
\draft
\preprint{YITP/K-1107}
\title{Transport coefficients, effective charge and mass for
multicomponent systems with fractional exclusion statistics}
\author{ Takahiro Fukui\footnote{JSPS Research Fellow}\footnote{e-mail
address:  fukui@yukawa.kyoto-u.ac.jp} and Norio Kawakami}
\address{Yukawa Institute for Theoretical Physics, Kyoto University,
Kyoto 606-01, Japan}
\date{April 1995}
\maketitle
\begin{abstract} 
Transport properties of the multicomponent
quantum many-body systems obeying Haldane's fractional
exclusion statistics are studied in one dimension.
By computing the finite-size spectrum
under twisted boundary conditions, we explicitly express
the conductivity and the conductance
in terms of statistical interactions. Through this analysis,
the effective charge and effective mass  for collective
excitations are determined. We apply the results for
$1/r^2$ quantum systems as well as  for correlated electron systems.
\end{abstract}
\section{Introduction}

In low-dimensional quantum systems, excitations are described by
quasiparticles carrying fractional quantum numbers. One of the well
known examples is the fractional quantum Hall effect (FQHE)\cite{FQHE},
where quasiparticles are classified by the
fractional charge and statistics. In these theories, fractional
quantum numbers arise from exchange properties of the wavefunction.
Recently, Haldane \cite{HAL} proposed a new
concept of fractional statistics based on the
state-counting of many-body systems, which is  a
generalization of Pauli-exclusion principle. We will refer to this as
fractional exclusion statistics. Thermodynamic properties were
already investigated in detail\cite{MURTHY,WU,NAYAK,FUKUI}.
For example, Wu and Bernard \cite{WU} formulated thermodynamic
equations, and  showed that the statistical interaction
is related with the two-body phase shift for Bethe-ansatz
solvable models. Their method was generalized
to multicomponent systems \cite{FUKUI}, and low-energy
critical properties were investigated.

In this paper, we study transport properties of the
multicomponent quantum systems with exclusion statistics in
one dimension. Transport coefficients  are closely related
with wavefunctions, and may usually be calculated
in the Green function formalism.
Since we have only statistics among particles without explicit
wavefunctions, it is not trivial to study such
properties directly from the definition of statistics.
We therefore use a trick to avoid this difficulty.
Namely, by applying the idea of twisted boundary conditions
for the finite-size spectrum, we calculate the conductivity
and the conductance
in terms of statistical interactions. We then determine
the effective charge and the effective mass, and show how these
quantities are related with exclusion statistics.

After a brief introduction
of exclusion statistics in the next section, we compute the
finite-size corrections due to the vector potential for
multicomponent quantum systems in section 3.
In section 4, we then obtain the conductivity and the
conductance, and determine the effective charge and
the effective mass. In section 5, we apply the results for
several interesting quantum systems. Section 6 is devoted to
summary and discussions.

\section{Exclusion statistics}

Let us start with a brief introduction for fundamental
properties of exclusion statistics\cite{HAL,WU}. It is based
on counting the change of the dimension of the one-particle Hilbert space
when a particle is added to the system, which is
explicitly formulated as,
\begin{equation}
\frac{\partial D_\alpha (k_\alpha )}{\partial N_\beta (k'_\beta )}
=- g_{\alpha\beta}(k_\alpha -k'_\beta )\equiv
-\{g'_{\alpha\beta}(k_\alpha -k'_\beta )+
\delta_{\alpha\beta}\delta_{k_\alpha k'_\beta }\},
\label{SI}
\end{equation}
where $D_\alpha (k_\alpha )$ and $N_\alpha (k_\alpha )$ are,
roughly speaking, the numbers of unoccupied (hole) and occupied
(particle) states specified by the internal quantum numbers $\alpha
=(1,2,...,M)$ and corresponding  momentum $k_\alpha$.
The matrix $g_{\alpha\beta}$, which is called
{\it statistical interaction}, describes correlation effects among
particles. For more rigorous definition, see Ref.\cite{HAL}.
Simple cases $g_{\alpha\beta}(k_\alpha -k'_\beta )=
g\delta_{\alpha\beta}\delta_{k_\alpha k'_\beta }$ with $g=1$ and $g=0$
correspond to free fermions and free bosons, respectively, and for
general fractional value $g$, we call
it {\it ideal fractional exclusion statistics}.

Statistical interactions should be independent
of $N_\alpha$\cite{HAL}, and hence eq.(\ref{SI}) results in
\begin{equation}
D_\alpha (k_\alpha )=-\sum_{\beta ,k'_\beta}g_{\alpha\beta}
(k_\alpha -k'_\beta )
N_\beta (k'_\beta )+D^0_\alpha (k_\alpha ).
\label{SII}
\end{equation}
We assume that integral constants are given
by $D^0_\alpha (k_\alpha )=
D^0\delta_{\alpha 1}$ or $D^0$, which are referred to
as hierarchical and symmetric bases, respectively.
Such bases are originally used for a classification
of the FQHE\cite{WEN,WENREV}. Also, in one-dimensional quantum
systems, the hierarchical basis serves as a natural basis for
the Bethe-ansatz solution\cite{YANG}.

In the thermodynamic limit, we introduce the
distribution functions for particles
and holes,
\begin{equation}
\rho_\alpha (k_\alpha )=\frac{N_\alpha (k_\alpha )}{D^0},\quad
\rho_\alpha^{(h)} (k_\alpha )=\frac{D_\alpha (k_\alpha )}{D^0},
\label{DFUN}
\end{equation}
where $D^0$ is proportional to the system size $L$ such
as $D^0=L/2\pi$ under periodic boundary conditions\cite{WU}.
The bare charge for each elementary excitation
is defined by\cite{FUKUI}
\begin{equation}
t_\alpha =\frac{D_\alpha^0}{D^0}.
\end{equation}
Consequently, eq.(\ref{SII}) can be written as
\begin{equation}
\rho_\alpha (k_\alpha )+
\rho_\alpha^{(h)}(k_\alpha )=t_\alpha -\sum_\beta
\int_{-\infty}^\infty
dk'_\beta g'_{\alpha\beta}(k_\alpha -k'_\beta )
\rho_\beta (k'_\beta ),
\label{BAE}
\end{equation}
The energy of the system is assumed
to take the form\cite{WU}:
\begin{equation}
\varepsilon\equiv\frac{E}{D^0}=\sum_\alpha\int_{-\infty}^{\infty}dk_\alpha
\epsilon_\alpha^0(k_\alpha )\rho_\alpha (k_\alpha )
\end{equation}
with the bare energy function $\epsilon_\alpha^0(k)$.
Note that many-body effects due to exclusion
statistics are incorporated in the
distribution function $\rho(k)$.

Thermodynamic equations are generally
obtained in a set of coupled nonlinear
equations at finite temperatures.\footnote{
Thermodynamics of systems with exclusion
statistics can be formulated by the method proposed in
Ref.\cite{WU}, and a multicomponent generalization
can be found in Ref.\cite{FUKUI}.
A key idea is to introduce the entropy $S=\ln W$ with
$$
W=\prod_{\alpha ,k_\alpha}\frac{(D_\alpha +N_\alpha -1)!}{N_\alpha !
(D_\alpha -1)!},
$$
which plausibly interpolates the boson and fermion cases\cite{HAL,WU}.
Note that if we choose $g_{\alpha\beta}$ for free fermions and
bosons, $\rho$ in eq.(\ref{DFUN})
reduces to the Fermi and Bose distribution functions.}
{}~Here we restrict ourselves to the zero-temperature
case which is sufficient for the  following calculation.
At zero temperature without external fields, $M$ species of
elementary excitations are specified by
the dressed energy function $\epsilon_\alpha(k)$.
The ``Fermi level" $Q_\alpha$ for each excitation
is determined by the conditions,
$\epsilon_\alpha (k_\alpha )<0$ for $|k_\alpha |<Q_\alpha$ and
$\epsilon_\alpha (k_\alpha )>0$ for $|k_\alpha |>Q_\alpha$,
provided that the energy dispersion may be symmetric
around the origin $k_{\alpha}=0$.
Then, eq.(\ref{BAE}) reduces to the following integral equations
supplemented by those for the dressed energy $\epsilon_{\alpha}$,
\begin{eqnarray}
&&\rho_\alpha (k_\alpha )=t_\alpha -\sum_\beta\int_{-Q_\beta}^{Q_\beta}
dk'_\beta g'_{\alpha\beta}(k_\alpha -k'_\beta )\rho_\beta (k'_\beta ),
\label{RHO}\\
&&\epsilon_\alpha (k_\alpha )=\epsilon_\alpha^0 (k_\alpha )-\mu t_\alpha
-\sum_\beta\int_{-Q_\beta}^{Q_\beta}
dk'_\beta g'_{\beta\alpha}(k'_\beta -k_\alpha)\epsilon_\beta
(k'_\beta ).
\label{BASIC}
\end{eqnarray}
The total energy is now expressed by the dressed energy as
\begin{equation}
\varepsilon =\sum_\alpha \varepsilon_\alpha +\mu n_c ,
\quad \varepsilon_\alpha =t_\alpha\int_{-Q_\alpha}
^{Q_\alpha}dk_\alpha\epsilon_\alpha (k_\alpha ),
\end{equation}
where  $\mu$ is the chemical potential and $n_c$
is the density of charged particles.
These equations can describe static properties
at zero temperature.

\section{Finite-size corrections due to static vector potential}

We now turn to the computation of the conductivity
in pure systems without randomness, which
can be calculated with the response to the vector
potential. Consider the ring system at $T=0$ threaded
by the magnetic flux, which gives rise to static vector potential
along the ring\cite{KOHN}. The effect of the vector potential
is incorporated in twisted boundary conditions\cite{SHASTRY,KAWAKAMI}.
Therefore, the energy increment quadratically proportional to $A$
can be calculated through the analysis of the finite-size spectrum,
which directly gives the charge stiffness and hence the
conductivity.  By observing that the basic equations
(\ref{RHO}) and (\ref{BASIC}) have the same structure
as the Bethe-ansatz equations, we can apply elegant
techniques of the dressed charge matrix
developed for integrable models\cite{IZERGIN},
and generalize the calculation of the
conductivity\cite{SHASTRY,KAWAKAMI} to multicomponent cases.

Before proceeding with the finite-size corrections,
we here give a formal solution to (\ref{RHO}) in the absence
of external fields, which is necessary for the
following discussions. To this end, let us
first introduce the functions\cite{IZERGIN,KOREPIN}
\begin{eqnarray}
&&K_{\alpha\beta}(k_\alpha -k'_\beta )=g'_{\alpha\beta}
(k_\alpha -k'_\beta )
-\sum_\gamma\int_{-Q_\gamma}^{Q_\gamma}dk''_\gamma
g'_{\alpha\gamma}(k_\alpha -k''_\gamma )K_{\gamma\beta}
(k''_\gamma -k'_\beta ) ,\label{DEFK}\\
&&Z_{\alpha\beta}(k_\beta )=\delta_{\alpha\beta}
-\sum_\gamma\int_{-Q_\gamma}^{Q_\gamma}dk'_\gamma
Z_{\alpha\gamma}(k'_\gamma )g'_{\gamma\beta}
(k'_\gamma -k_\beta ),
\label{DEFDC}
\end{eqnarray}
where $Z_{\alpha\beta}$ is called the dressed charge
matrix \cite{IZERGIN}. By noting the relation
\begin{equation}
Z_{\alpha\beta}(k_\beta )=\delta_{\alpha\beta}-\int_{-Q_\alpha}
^{Q_\alpha}dk'_\alpha K_{\alpha\beta}(k_\alpha -k'_\beta),
\end{equation}
we can formally write down the
distribution function in terms of the dressed charge matrix,
\begin{equation}
\rho_\alpha (k_\alpha )=\sum_\beta t_\beta Z_{\beta\alpha}(k_\alpha ).
\label{RZ}
\end{equation}

We now evaluate finite-size corrections to
the total energy due to the vector potential, which is the
same as those for twisted boundary
conditions\cite{SHASTRY,KAWAKAMI}. First, by observing that
the effects of  the static vector potential $A$ are to shift the
momentum  by the amount $\delta_\alpha$ proportionally
to $A$ (in unit $e$), the basic equations should read,
\begin{eqnarray}
&&\widetilde\rho_\alpha (k_\alpha )=t_\alpha
-\sum_\beta\int_{-Q_\beta +\delta_\beta}^{Q_\beta +\delta_\beta}
dk'_\beta g'_{\alpha\beta}(k_\alpha -k'_\beta )
\widetilde\rho_\beta (k'_\beta ) ,\\
&&\widetilde\epsilon_\alpha (k_\alpha )
=\epsilon_\alpha^0 (k_\alpha )-\mu t_\alpha
-\sum_\beta\int_{-Q_\beta +\delta_\beta}^{Q_\beta +\delta_\beta}
dk'_\beta g'_{\beta\alpha}(k'_\beta -k_\alpha)\widetilde\epsilon_\beta
(k'_\beta )\label{TILDEPS} ,\\
&&\widetilde \varepsilon =\sum_\alpha\int_{-Q_\alpha +\delta_\alpha}
^{Q_\alpha +\delta_\alpha}
dk_\alpha\epsilon^0(k_\alpha )\widetilde\rho_\alpha (k_\alpha )
=\sum_\alpha \widetilde\varepsilon_\alpha +\mu n_c ,
\label{EE}
\end{eqnarray}
where
\begin{equation}
\widetilde\varepsilon_\alpha =t_\alpha\int_{-Q_\alpha +\delta_\alpha}
^{Q_\alpha +\delta_\alpha}
dk_\alpha\widetilde\epsilon (k_\alpha ).
\end{equation}
It is to be noted that the  vector potential
not only shifts the  momentum uniformly,
but also can
rearrange in general the distribution of the momentum
via interactions among particles.  For clarity,
we put tilde for rearranged quantities.

Let us compute the corrections to the total energy
(\ref{EE}). By differentiating eq.(\ref{TILDEPS})
with respect to $\delta$, we have
\begin{equation}
\left.\frac{\partial\widetilde\epsilon_\alpha (k_\alpha )}
{\partial\delta_\beta}\right|_{\delta =0}=0.
\end{equation}
by using the conditional equation for the ``Fermi level",
$\epsilon_\alpha (Q_\alpha )=0$.
By differentiating again, we have
\begin{eqnarray}
\left.\frac{\partial^2\widetilde\epsilon_\alpha (k_\alpha )}
{\partial\delta_\beta\partial\delta_\gamma}\right|_{\delta =0}
&=&-\epsilon '_\beta (Q_\beta )\{g'_{\beta\alpha}(Q_\beta -k_\alpha )
+g'_{\beta\alpha}(-Q_\beta -k_\alpha )\}\delta_{\beta\gamma}
\nonumber\nonumber\\
&&-\sum_\gamma\int_{-Q_\gamma}
^{Q_\gamma}dk'_\gamma g'_{\alpha\gamma}(k_\alpha -k'_\gamma )
\left.\frac{\partial^2\widetilde\epsilon_\gamma (k'_\gamma )}
{\partial\delta_\beta\partial\delta_\gamma}\right|_{\delta =0},\nonumber
\end{eqnarray}
which can be formally
solved by the iteration scheme, resulting in
\begin{equation}
\left.\frac{\partial^2\widetilde\epsilon_\alpha (k_\alpha )}
{\partial\delta_\beta\partial\delta_\gamma}\right|_{\delta =0}
=-\epsilon '_\beta (Q_\beta )\{K_{\beta\alpha}(Q_\beta -k_\alpha )
+K_{\beta\alpha}(-Q_\beta -k_\alpha )\}\delta_{\beta\gamma},
\end{equation}
where prime in $\epsilon '_\alpha$ stands for the derivative with respect
to $k_\alpha$.
In the following, we assume the relation $g_{\alpha\beta}(k_\alpha
-k_\beta ')=g_{\beta\alpha}(k_\beta '-k_\alpha )$.
Using these results, we have
\begin{eqnarray}
\left.\frac{\partial\widetilde\varepsilon_\alpha}{\partial\delta_\beta}
\right|_{\delta =0}&=&0,\\
\left.\frac{\partial^2\widetilde\varepsilon_\alpha}
{\partial\delta_\beta\partial\delta_\gamma}\right|_{\delta =0}&=&
t_\alpha\epsilon '_\beta (Q_\beta )\left[2\delta_{\alpha\beta}-
\int_{-Q_\alpha}^{Q_\alpha}dk_\alpha\{K_{\beta\alpha}(Q_\beta -
k_\alpha )+K_{\beta\alpha}(-Q_\beta -k_\alpha
)\}\right]\delta_{\beta\gamma}\\
&=&2t_\alpha\epsilon '_\beta (Q_\beta ){\cal Z}_{\alpha\beta}
\delta_{\beta\gamma},
\label{DERIV}
\end{eqnarray}
where ${\cal Z}_{\alpha\beta}\equiv Z_{\alpha\beta}(Q_\beta)$
is the dressed charge matrix defined in (\ref{DEFDC}).
Therefore, we obtain the finite size correction $\Delta
\varepsilon\equiv\widetilde\varepsilon -
\varepsilon$ due to the small shift of $\delta_\alpha$ as,
\begin{eqnarray}
\Delta\varepsilon&=&\frac{1}{2}\sum_{\alpha\beta\gamma}
\delta_\beta\left.\frac{\partial^2\widetilde\varepsilon_\alpha}
{\partial\delta_\beta\partial\delta_\gamma}\right|_{\delta =0}
\delta_\gamma \nonumber\\
&=&\sum_{\alpha\beta}t_\alpha ({\cal Z}V{\cal Z}^T)_{\alpha\beta}t_\beta
\delta_\beta\delta_\gamma ,
\label{FSCA}
\end{eqnarray}
where $V_{\alpha\beta}=v_\alpha\delta_{\alpha\beta}$, and $v_\alpha$
is the Fermi velocity defined by $v_\alpha =\epsilon'_\alpha (Q_\alpha
)/\rho_\alpha (Q_\alpha )$.
To derive the second line, we have used eqs.(\ref{RZ})
and (\ref{DERIV}).

The remaining task is to determine the relation between
$\delta_\alpha$ and $A$.  The result is quite simple (see eq.
(\ref{RELATE})), but still nontrivial,
which depends on types of external fields.
Here we briefly depict how to obtain this
relation in a bit general way\cite{KOREPIN}.
For this purpose, let us first introduce
\begin{eqnarray}
&&i_\alpha (k_\alpha )=\int^{k_\alpha}dk\rho_\alpha
(k), \quad
\widetilde i_\alpha (k_\alpha )=\int^{k_\alpha}dk
\widetilde\rho_\alpha (k),\\
&& G_{\alpha\beta}(k_\alpha -k'_\beta )=
\int^{k_\alpha}dkg'_{\alpha\beta}(k-k'_\beta ).
\end{eqnarray}
When $A=0$, eq.(\ref{RHO}) is integrated as
\begin{equation}
i_\alpha (k_\alpha )=t_\alpha k_\alpha -
\sum_\beta\int_{-Q_\beta}^{Q_\beta}
dk'_\beta G_{\alpha\beta}(k_\alpha -k'_\beta )
\rho_\beta (k'_\beta ).
\label{eq:IGS}
\end{equation}
Notice that the first term in the
right hand side is regarded as a bare momentum of the
system, which can be written as $p_\alpha^0\equiv
t_\alpha k_\alpha$.
In the presence of the vector potential, the momentum is shifted
as $p_\alpha^0\rightarrow p_\alpha^0-At_\alpha$. Then
eq.(\ref{eq:IGS}) is modified to
\begin{equation}
\widetilde i_\alpha (k_\alpha )+d_\alpha =t_\alpha k_\alpha -
\sum_\beta\int_{-Q_\beta +\delta_\beta}^{Q_\beta +\delta_\beta}
dk'_\beta G_{\alpha\beta}(k_\alpha -
k'_\beta )\widetilde\rho_\beta (k'_\beta ),
\label{eq:IEF}
\end{equation}
where we denote $d_\alpha \equiv At_\alpha$.
Note that the following
calculation can also be applied for other types of external fields
if we suitably take $d_\alpha$ different from $At_\alpha$.
Now introduce $\widetilde k_\alpha =\widetilde k_\alpha (k_\alpha )$
such that $i_\alpha (k_\alpha )=\widetilde
i_\alpha (\widetilde k_\alpha )$, then we find
$\widetilde k_\alpha (Q_\alpha )=Q_\alpha +\delta_\alpha$ and
\begin{equation}
\rho_\alpha (k_\alpha )dk_\alpha =\widetilde\rho_\alpha (\widetilde
k_\alpha )d\widetilde k_\alpha .
\end{equation}
By the use of these relations, we can change the integral variables
into $\tilde k_\alpha$ in eq.(\ref{eq:IEF}).
By subtracting both sides of eq.(\ref{eq:IGS}) from (\ref{eq:IEF})
and expanding $G_{\alpha\beta}(\widetilde k_\alpha -
\widetilde k'_\beta )$ up to the first order in $(\widetilde k-k)$,
we get
\begin{equation}
F_\alpha (k_\alpha )=d_\alpha-\sum_\beta\int_{-Q_\beta}^{Q_\beta}
dk'_\beta g'_{\alpha\beta}(k_\alpha -k'_\beta )F_\beta (k'_\beta ),
\label{F}
\end{equation}
where $F_\alpha$ is defined by\cite{KOREPIN}
\begin{equation}
F_\alpha (k_\alpha )=(\widetilde k_\alpha -k_\alpha )
\rho_\alpha (k_\alpha).
\label{DEFF}
\end{equation}
The formal solution to eq.(\ref{F}) is given by
\begin{equation}
F_\alpha (k_\alpha )=\sum_\beta d_\beta Z_{\beta\alpha}(k_\alpha ).
\label{SOLF}
\end{equation}
As stressed above,
this formula is valid for arbitrary types of $d_\alpha$. By explicitly
substituting $d_\alpha \equiv At_\alpha$, and comparing (\ref{DEFF})
and (\ref{SOLF}) by the use of (\ref{RZ}), we
get the simple relation,
\begin{equation}
\widetilde k_\alpha -k_\alpha =A=\delta_\alpha .
\label{RELATE}
\end{equation}
Namely, momentum shifts occur not only for the charge sector but
for all indices $\alpha$.

Consequently, by combining (\ref{FSCA}) and (\ref{RELATE}),
  we end up with the final formula for
the energy increment as,
\begin{equation}
\Delta\varepsilon
=\sum_{\alpha\beta}t_\alpha ({\cal Z}V{\cal Z}^T)_{\alpha
\beta}t_\beta A^2.
\label{FSC}
\end{equation}
This completes the calculation of finite-size corrections
due to the static vector potential.


\section{Transport coefficients, effective charge and mass}
\subsection{conductivity}

According to the formula (\ref{FSC}) for the
response to the external vector potential,
we can obtain the charge stiffness as,
\begin{equation}
D_c =\sum_{\alpha\beta}t_\alpha ({\cal Z}V{\cal
Z}^T)_{\alpha\beta}t_\beta .
\label{CS}
\end{equation}
An important point is that $D_c$ is directly related
with the conductivity through the relation,
\begin{equation}
{\rm Re} \, \sigma (\omega)=e^2D_c\delta (\omega)
\quad {\rm at}\quad\omega =0
\end{equation}
according to the linear response theory\cite{KOHN,SHASTRY}.

In many literatures so far, correlation effects on $D_c$
have been  considered to modify only the effective
transport mass $m^*$\cite{KOHN,SHASTRY,KAWAKAMI}.
However, the transport mass
is not sufficient to describe the correlation effects
on transport properties, because quasiparticles
such as holons in one dimension can carry not
only the effective mass but
also the effective charge. Therefore, in place of
ordinary interpretation\cite{SHASTRY,KAWAKAMI},
we propose the following natural
expression for the conductivity in terms of
the effective charge $e^*$ and effective mass $m^*$,
\begin{equation}
{\rm Re}\sigma (w)=\frac{\pi ee^*n_c}{m^*}\delta (\omega )
\quad{\rm at}\quad\omega =0 ,
\label{RECON}
\end{equation}
where $n_c$ is the density of charged particles.
The charge stiffness calculated in eq.(\ref{CS})
is then related with $D_c$ such that
\begin{equation}
D_c\propto\frac{e^*}{m^*}.
\label{EFFECTM}
\end{equation}
So, the charge stiffness (or conductivity) is not sufficient
to derive the effective charge and mass, and another
quantity is necessary to determine them.
We will show that the effective charge can be
derived from the conductance.

\subsection{conductance}

It is known that the fractional charge enters in the
conductance for the finite system in one dimension.
Though it is not easy to calculate the conductance
without wavefunctions, we can compute it
by taking into account a universal property of
Luttinger liquids, i.e., {\it the conductance is solely
determined by the correlation exponent for the charge sector}.
To this end, we first define the critical exponent
for the charge density correlation function in the asymptotic region,
\begin{equation}
\langle\rho (x)\rho (0)\rangle
\sim\exp (2ik_F^{(c)} x)x^{-2M f_c},
\end{equation}
where $k_F^{(c)}=\pi n_c$ is the ``Fermi momentum" for the
charge sector, and is usually given by $k_F^{(c)}=M k_F$
in terms of the ordinary Fermi momentum $k_F$.
The correlation exponent  $f_c$
is a function of statistical interactions,
which is normalized to reproduce
$f_c =1$ for non-interacting systems.
An important point is that the conductance $G_c$ can be
determined by $f_c$ in a universal way\cite{MESO},
\begin{equation}
G_c=M\frac{e^2}{h}f_c .
\label{COND}
\end{equation}
This formula can be deduced by observing that the conductance
is controlled only by the charge degrees of
freedom.\footnote{Exponent $g$ without randomness in ref.\cite{MESO}
corresponds to $f_c$ in this paper.}
Renormalizing $e$ by $f_c$ such that $G_c=Mee^*/h$,
we can  naturally define the effective charge as
\begin{equation}
e^*=f_c e.
\label{EFFECTC}
\end{equation}
Therefore, the remaining task is to obtain the critical exponent
$f_c$ in terms of statistical interactions.
Following a way similar to the last section, we can  derive
the exponent $f_c$ through conformal-field-theory analysis
of the finite-size spectrum which has been already
computed in \cite{FUKUI}.  We thus obtain the
renormalizarion factor for the fractional charge,
\begin{equation}
f_c =\frac{1}{M}\sum_{\alpha\beta}t_\alpha ({\cal
ZZ}^T)_{\alpha\beta}t_\beta ,
\label{RENORM}
\end{equation}
in terms of statistical interactions which are
implicitly  incorporated
in the dressed charge matrix. By applying the above formulas
to eq.(\ref{EFFECTM}), we can also extract the effective
transport mass $m^*$, which turns out to be inversely
proportional to the velocity $v$. The expressions
(\ref{CS}), (\ref{EFFECTM}), (\ref{EFFECTC})
and (\ref{RENORM}) are the  main results in this paper.

One can see that the effective charge
is determined {\it solely by the statistical interactions},
whereas the effective mass also depends on
non-universal quantities such as the velocity.  We wish to
emphasize that the formula for the fractional
charge (\ref{EFFECTC}) with (\ref{RENORM}) is universal, which
holds generally  for multicomponent Luttinger liquids.


\section{Applications}

\subsection{Ideal fractional exclusion statistics}

One of the most remarkable applications of
exclusion statistics is that for the {ideal} case,
in which the statistical interaction
is given in a simple form\cite{HAL,MURTHY,WU,NAYAK,FUKUI},
\begin{equation}
g_{\alpha\beta}(k_\alpha -k'_\beta )=G_{\alpha\beta}\delta
(k_\alpha -k'_\beta ) .
\label{IDEAL}
\end{equation}
This model is known to have close relationship
to interesting quantum systems such as
the FQHE, the $1/r^2$ systems, etc.

If we take a bare dispersion
as $\epsilon_\alpha^0 (k)=\epsilon^0(k)
t_\alpha$ with $\epsilon (k)=k^2/2$, the
ground state configuration is $Q_1=Q_2=...=Q_M\equiv Q$,
and the distribution functions are
obtained as
\begin{equation}
\rho_\alpha =\sum_\beta G_{\alpha\beta}^{-1}t_\beta .
\end{equation}
The number of charged particles is 
\begin{equation}
n_c=2Q\nu ,\quad \nu =\sum_{\alpha\beta}t_\alpha G^{-1}_{\alpha\beta}
t_\beta,
\label{NU}
\end{equation}
where the quantity $\nu$ is related to the compressibility as
$\kappa_c=4\nu^{2}/n_c$, which is a one-dimensional analogue
of the filling factor in the FQHE.
In the present model, the Fermi velocities for each excitation
take the same values
\begin{equation}
v_1=v_2=...=v_M\equiv v={n_c \over 2\nu}.
\label{FILLINGDEF}
\end{equation}
As is shown in Appendix, one finds a simple relation
between the dressed charge and the statistical interaction,
\begin{equation}
{\cal ZZ}^T=G^{-1}.
\label{ZZG}
\end{equation}
Consequently, the renormalization factor (\ref{RENORM})
for the fractional charge $e^*/e=f_c$
is expressed as
\begin{equation}
f_c =\frac{1}{M}\sum_{\alpha\beta}t_\alpha (
G^{-1})_{\alpha\beta}t_\beta .
\label{FFF}
\end{equation}
It is to be noted that the fractional charge is
now explicitly  obtained  only in terms of the
statistical parameters $G_{\alpha\beta}$ for exclusion statistics.
Since the charge stiffness is given by $D_c=n_c/2$,
the enhancement factor for the transport mass in this case is
derived
\begin{equation}
m^*/m =f_c,
\label{MMM}
\end{equation}
from which one can see that the enhancement of $m^*$
exactly cancels the renormalization of charge $e^*$, in accordance with
the translational symmetry.

Let us now discuss more concrete models for ideal exclusion
statistics. For an instructive example, we
consider the statistical-interaction matrix $G$
in hierarchical basis, which is derived from
the continuum $1/r^2$
model\cite{ISMD} with SU($M$) symmetry\cite{ISMS,ISMM}.
This model is also related to
a fundamental series for the hierarchical
FQHE\cite{WEN,BLOK,READ}. The $M \times M$ matrix
for statistical interactions in this case reads,
\begin{equation}
G=\left(\begin{array}{ccccc}2n+1&-1&0&&\\-1&2&-1&0&\\
&&\ddots&&\\&0&-1&2&-1\\&&0&-1&2\end{array}\right).
\label{G}
\end{equation}
We have from eq.(\ref{NU}),
\begin{equation}
\nu =\frac{1}
{2n+1-\displaystyle{\frac{1}{2-\displaystyle{\frac{1}{\ddots
\displaystyle{\frac{1}{2}}}}}}}=\frac{M}{2Mn+1}\equiv\frac{p}{q},
\label{FILLING}
\end{equation}
which corresponds to the filling factor in the case of the FQHE.
Observing this, we see that the matrix $G$ in
this model plays a role similar to the flux attachment
in Jain's model for the FQHE\cite{JAIN};
\begin{equation}
G\quad\longleftrightarrow\quad\chi_M\chi_1^{2n},
\end{equation}
where $\chi_M$ is the IQH state with the filling factor $M$,
which is attached by $2n$ flux quanta $\chi_1^{2n}$.
In fact,  one can find the same matrix as
(\ref{G}) in the classification scheme in the corresponding
Abelian Chern-Simons theory for the FQHE\cite{WEN,WENREV,BLOK,READ}.
Therefore, it is seen  that the SU($M$) $1/r^2$ model
has close relationship to the hierarchical
FQHE with $\nu=M/(2Mn+1)$.

According to eqs.(\ref{FFF}) and (\ref{MMM}), it turns out
that the effective charge and mass are given by
taking ${\bf t}^T=(1,0, \cdots, 0)$,
\begin{eqnarray}
&&e^*=\frac{e}{2Mn+1}=\frac{e}{q},\label{ISMC}\\
&&m^*=\frac{m}{2Mn+1}=\frac{m}{q}.
\label{ISMM}
\end{eqnarray}
Note that the expression for the effective charge (\ref{ISMC})
is actually in accordance  with that for the FQHE.
In particular,  for one component case $G=g$, we have
$e^*/e=m^*/m=1/g$.
Note that for free systems, i.e., $n=0$, we have $e^*/e=m^*/m=1$.

\subsection{Correlated electron systems}

It is also instructive to apply the results
to one-dimensional correlated electron systems.
In order to fully describe interacting electron systems
in the whole energy range, it is necessary to consider
statistical interactions depending on the momentum in a
complicated way\cite{HATSUGAI}.
However, if we restrict ourselves to the low-energy
conformal limit, we can still use the idea of
{\it ideal exclusion statistics}.
In such low-energy region, the critical behavior
is described by the Luttinger liquid theory,
in other words, by $c=1$ conformal field theory.
In this case, we can introduce $2 \times 2$
matrix $G_{\alpha\beta}$  for ideal statistics
in (\ref{IDEAL}) in terms of Luttinger liquid parameters.
Since this model has two kinds of elementary excitations, i.e., spinon
and holon which have two different velocities,
$v_s$ and $v_c$, we can choose the hierarchical basis as a natural one.
By analyzing exactly solved models
or Tomonaga-Luttinger models, the matrix for statistical interaction
can be deduced as\cite{FUKUI},
\begin{equation}
G=\left(\begin{array}{cc}\displaystyle{\frac{K_\rho +1}{2K_\rho}}
&\quad -1\\-1&2\\
\end{array}\right) ,
\label{EL}
\end{equation}
where $K_\rho={\cal Z}_{11}^2/2$
is the critical exponent for the
$4k_F$ oscillation piece in the density correlation function.
Note that $G_{11}$ is related with the charge degrees of freedom,
$G_{22}$ to the spin degrees of freedom, and the off-diagonal
elements are regarded as mutual statistics.
It is to be emphasized here  that (\ref{EL}) is the universal
formula for correlated electron systems.


According to eq.(\ref{EFFECTC}), the effective charge
is given by substituting ${\bf t}^T=(1,0)$,
\begin{equation}
e^*/e=K_\rho,
\label{TJC}
\end{equation}
which reproduces the known results
for the conductance in Luttinger liquids\cite{MESO}.
Also, from eq.(\ref{CS}), the charge stiffness is found to be
\begin{equation}
D_c=2v_c K_{\rho} ,
\end{equation}
from which we obtain the effective mass using eq.(\ref{EFFECTM}),
\begin{equation}
m^*/m=v_F/ v_c .
\end{equation}
where $v_F$ is the Fermi velocity for
non-interacting electrons.

For exactly solvable electron models such as the
Hubbard model and the supersymmetric {\it t-J}
model, the critical exponent $K_\rho$ and the
velocity of holons were calculated
exactly as functions of the interaction and the electron density
\cite{SCHLOTTMANN,KAWAKAMIYANG,HUB}.
So, we can discuss the effective charge and mass for these models.
As for the Hubbard model, $K_\rho$ decreases
from 1 to 1/2 as the Coulomb interaction is increased\cite{HUB},
hence the effective charge decreases with the increase of the
interaction,  as stressed by Ogata and Fukuyama\cite{MESO}.
 Near half filling, the effective charge always takes
$e/2$ as far as the Coulomb interaction exists.
In the case of the supersymmetric {\it t-J} model
\cite{SCHLOTTMANN,KAWAKAMIYANG},
the effective charge is $e/2$ near half filling, but
as the electron density decreases, it continuously
increases and reaches the non-interacting value $e^*=e$
in the low density limit.  Also, we can discuss the
effective mass for electron systems. The results are
essentially same as previously discussed\cite{KAWAKAMI}:
the effective mass has a divergence property near half filling
both for the Hubbard model and the supersymmetric {\it t-J}
model, reflecting the metal-insulator transition.

\section{summary and discussions}

In summary, we have obtained the transport coefficients,
the  effective charge and mass
for multicomponent quantum systems obeying fractional
exclusion statistics. Their explicit relation to
the statistical interaction has been derived
in eqs.(\ref{CS})$\sim$ (\ref{RENORM})
for general systems obeying (\ref{SI}).
We have applied the results for the cases with ideal statistics as
well as for the conformal limit of electron systems.
It  has been  also pointed out that
the statistical interaction derived from SU($N$) $1/r^2$ models
are closely related with Jain's construction (or the
corresponding Chern-Simons theory) for the
hierarchical FQHE.

It is instructive to note that the effective charge
in the {\it ideal} case is expressed in an extremely simple form
(\ref{FFF}) or (\ref{ISMC}) in terms of the statistical
interaction $G_{\alpha\beta}$, implying that the
fractional charge in the ideal case directly
reflects the fractional statistics.
In fact, we find an alternative way to derive
(\ref{ISMC}) using only
the definition of fractional exclusion statistics.
We briefly summarize how to get them intuitively.
In the ideal case, the definition eq.(\ref{SI}) reads
\begin{equation}
\frac{\partial N_\alpha}{\partial D_\beta}=-G_{\alpha\beta}^{-1}
\quad {\rm at}\quad k_\alpha =k_\beta .
\end{equation}
Now imagine the ground state configuration and
make a hole to excite the system. The above equation implies that
if we make a $\beta$-hole, the number of
$\alpha$-particles decreases by the amount
of $G_{\alpha\beta}^{-1}$. Then, how many charged particles decrease
in all? The answer is
$
\sum_\alpha t_\alpha G_{\alpha\beta}^{-1}.
$
Now make a hole at a sector
$\alpha$, i.e., ${\bf t}^T=(0,...,1,0,...0)$
in the hierarchical basis. This corresponds to ${\bf t}^T
=(0,...,1,1,...,1)$
in the symmetric basis, which may create $M+1-\alpha$ holes of
electrons. Therefore, making a hole with unit charge corresponds to
removing particles with the charge
\begin{equation}
e_\alpha^*=\frac{e}{M+1-\alpha}\sum_\beta G_{\beta\alpha}^{-1},
\label{EACHC}
\end{equation}
which in turn defines the effective charge
of the excitation.
If we adopt $G$ in eq.(\ref{G}), then we have
\begin{equation}
e_\alpha^*/e=1/q ,
\label{EXPLICITEACHC}
\end{equation}
where $q$ is defined in eq.(\ref{FILLING}).
This result coincides with (\ref{ISMC}).


Finally, another remark is in order. In section III, we have
defined the effective charge (\ref{EFFECTC}) apart from the trivial
degeneracy $M$. However, it may be possible to include such factor in
the definition of the charge. In this definition, eqs.(\ref{FFF}),
(\ref{ISMC}) and (\ref{ISMM}) are modified by the factor $M$,
and eq.(\ref{EACHC}) should be replaced by
\begin{equation}
\widetilde e_\alpha^*=e\sum_\beta G_{\beta\alpha}^{-1}
\end{equation}
and therefore, eq.(\ref{EXPLICITEACHC}) is modified as
\begin{equation}
\widetilde e_\alpha^*/e=(M+1-\alpha )/q,
\end{equation}
i.e., explicitly, $p/q, (p-1)/q, ... ,1/q$.
This definition for the fractional charge  corresponds to that
used  in refs.\cite{WEN,WENREV}.

\acknowledgements

This work is supported by the Grant-in-Aid from the Ministry of
Education, Science and Culture.

\appendix
\section{}

In the case of ideal exclusion statistics,
there exists  a simpler way
to derive the formula (\ref{ZZG}) without
calculating  the dressed charge (\ref{DEFDC}).
We briefly depict this convenient method below.
Consider the present system without external fields.
Then there are two kinds of elementary excitation, i.e.,
the excitation which changes the
number of particles, and which carries the large momentum.
Following techniques of the dressed charge matrix\cite{IZERGIN},
the excitation spectrum was explicitly evaluated in \cite{FUKUI},
$$\Delta\varepsilon =(v/4){\bf n}^T({\cal ZZ}^T)^{-1}{\bf n}+
v{\bf d}^T({\cal ZZ}^T){\bf d},$$
where the vectors ${\bf n}$ and ${\bf d}$ denote
the  quantum numbers for excitations, which label the
change of particle number and the momentum transfer,
respectively. Note that the
excitation which we seek for in (\ref{FSC})
corresponds to the excitation specified by ${\bf d}$.
{}From the above formula, one can see that these
 two kinds of excitations
are related to each other reflecting modular invariance.
So, we can easily deduce the excitations for the ${\bf d}$-sector
once we can calculate those for the ${\bf n}$-sector.
The calculation for the latter excitation is much simpler
than the former. Let us then calculate the latter
by changing  $Q_\alpha\rightarrow Q_\alpha
+\Delta Q_\alpha$. Then both $n_\alpha$
and $\Delta\varepsilon$ are given by
functions of $\Delta Q_\alpha$, and a simple calculation
gives $\Delta\varepsilon$ as a function of $n_\alpha$
such that $\Delta\varepsilon =(v/4){\bf n}^TG{\bf n}$.
Comparing these results, we end up with  the formula (\ref{FFF}).


\end{document}